\documentclass[]{aa}

\usepackage[version=3]{mhchem}
\usepackage{threeparttable, tablefootnote}
\usepackage{xcolor}
\usepackage{multicol}

\usepackage{graphicx}
\usepackage{txfonts}

\begin{document} 

\title{Sulphur monoxide emission tracing an embedded planet in the HD~100546 protoplanetary disk}

\author{Alice S. Booth \inst{1},
John D. Ilee \inst{2},
Catherine Walsh \inst{2},
Mihkel Kama \inst{3,4},
Luke Keyte \inst{3},
Ewine F. van Dishoeck \inst{1,5},
Hideko Nomura \inst{6}
}

\institute{Leiden Observatory, Leiden University, 2300 RA Leiden, the Netherlands \\
\email{abooth@strw.leidenuniv.nl} 
\and 
School of Physics and Astronomy, University of Leeds, Leeds LS2 9JT, UK
\and
Department of Physics and Astronomy, University College London, Gower Street, London, WC1E 6BT, UK
\and
Tartu Observatory, University of Tartu, Observatooriumi 1, 61602 T\~{o}ravere, Tartumaa, Estonia
\and
Max-Planck-Institut f\"{u}r Extraterrestrishe Physik, Gie{\ss}enbachstrasse 1, 85748 Garching, Germany
\and
National Astronomical Observatory of Japan, 2--21--1 Osawa, Mitaka, Tokyo 181--8588, Japan
}

\titlerunning{Sulphur Monoxide in HD~100546}
\authorrunning{A. S. Booth et al.}

\abstract
{Molecular line observations are powerful tracers of the physical and chemical conditions across the different evolutionary stages of star, disk and planet formation. Using the high angular resolution and unprecedented sensitivity of the Atacama Large Millimeter Array (ALMA) there is now a drive to detect small scale gas structures in protoplanetary disks that can be attributed directly to forming planets. We report high angular resolution ALMA Band 7 observations of sulphur monoxide (SO) in the nearby planet-hosting disk around Herbig star HD~100546. SO is rarely \color{black} detected \color{black} in evolved protoplanetary disks but in other environments it is most often utilised as a tracer of shocks. The SO emission from the HD~100546 disk is primarily originating from gas within the $\approx$20~au mm-dust cavity and shows a clear azimuthal brightness asymmetry of a factor of 2. 
In addition, we see a significant difference in the line profile shape when comparing these new Cycle 7 data to Cycle 0 data of the same SO transitions. We discuss the different physical/chemical mechanisms that could be responsible for this asymmetry and time variability including disk winds,  disk warps, and a shock triggered by a (forming) planet. We propose that the SO is enhanced in the cavity due to the presence of a giant planet. The SO asymmetry \color{black} complements \color{black} evidence for hot circumplanetary material around the giant planet HD~100546~c traced via CO ro-vibrational emission. This work sets the stage for further observational and modeling efforts to detect and understand the chemical imprint of a forming planet on its parent disk. 
}

\keywords{Astrochemistry, Protoplanetary disks, Submillimeter: planetary systems, Planet-disk interactions}

   \date{Received July 17, 2022; accepted October 24, 2022}
 
\maketitle
%

\section{Introduction}

Observations of protoplanetary disks with the Atacama Large Millimeter Array (ALMA) have revealed structures in the millimeter dust and molecular gas that are clear departures from radially smooth and azimuthally symmetric disks \citep[e.g.,][]{2013Sci...340.1199V, 2015ApJ...808L...3A, 2018ApJ...869L..41A, 2021AJ....162...99F, 2021ApJS..257....1O, 2021AJ....161...33V, 2021A&A...651L...5V, 2021A&A...651L...6B}. Although the exact mechanism(s) underlying the formation of these structures is still under debate, ongoing planet formation within disk gaps is the favoured scenario \citep[e.g.][]{2018ApJ...869L..47Z, 2019ApJ...871..107F, 2019NatAs...3.1109P, 2020MNRAS.499.2015T, 2020ApJ...888L...4T}. But, due to the complex nature of the coupled physics and chemistry in disks it is non-trivial to connect the structures observed in molecular line emission back to the presence and the properties of the potential forming planets.

Most often the detected substructures in the dust and gas are cavities and concentric rings, but the link between the structures seen in the dust and those seen in the array of available gas tracers is not always entirely clear as not all dust gaps coincide with molecular gaps \citep[e.g.,][]{2015A&A...579A.106V, 2016A&A...585A..58V, 2021ApJS..257....1O, 2021ApJS..257....3L, 2021arXiv211213859J}. This is in part because molecular rings, especially in tracers other than CO, can arise from different chemical processes in the disk, e.g., they depend on the disk UV field and degree of ionisation, and thus do not solely trace the bulk disk gas density structure \citep[e.g.,][]{2016ApJ...831..101B, 2017A&A...599A.101V, 2017A&A...606A.125S, 2018A&A...609A..93C, 2021A&A...646A...3L}. In a few specific cases chemical structures in the gas can be very clearly linked to structures in the dust. Most strikingly, in the IRS~48 disk multiple different molecules clearly follow the millimeter dust distribution. In IRS~48 there is a clear chemical asymmetry: where the sulphur monoxide (\ce{SO}), sulphur dioxide (\ce{SO_2}), nitric oxide (\ce{NO}), formaldehyde (\ce{H_2CO}), methanol (\ce{CH_3OH}), and dimethyl ether (\ce{CH_3OCH_3}) are all co-spatial with the highly asymmetric dust trap \citep{2021A&A...651L...5V, 2021A&A...651L...6B, 2022A&A...659A..29B}. The most likely chemical origin of these species is thermal sublimation of \ce{H_2O} and more complex organic ices at the irradiated dust cavity edge. The thermal sublimation of ices, also seen in the HD~100546 disk via the detection of gas-phase \ce{CH_3OH} \citep{2021NatAs...5..684B}, gives us a window to access the full volatile content in the disk available to form planetesimals and comets. 

There are other instances of enhancement in gas phase volatiles from the ice reservoir that are more subtle. The inward transport of icy pebbles due to radial drift has been shown to enhance the abundance of CO within the snowline \citep{2017MNRAS.469.3994B, 2018ApJ...864...78K,2020ApJ...899..134K, 2019MNRAS.487.3998B, 2019ApJ...883...98Z, 2020ApJ...891L..16Z, 2021ApJS..257....5Z} and similarly, enhance the abundance of \ce{H_2O} in the inner disk \citep{2020ApJ...903..124B}. 
Molecules can also be removed from the grains via non-thermal desorption, at radii beyond their snowlines. One tracer of this in particular is \ce{H_2CO} which in a number of disks shows an increase in column density at the edge of the millimeter dust disk
\citep{2015ApJ...809L..25L, 2017A&A...605A..21C, 2018ApJ...863..106K, 2020ApJ...890..142P, 2021ApJS..257....6G}.
Chemical asymmetries can also arise in warped disks where there is no clear azimuthal asymmetry in the outer dust disk but rather a misaligned inner disk \citep{2021MNRAS.505.4821Y}.  Here the differences in chemistry are driven by the azimuthal variation in disk temperature due to shadowing from the inner disk. 

Disks are therefore chemically diverse environments and furthermore, planet formation can indirectly affect the observable chemistry. In particular, the physics driving the dust evolution, i.e., grain growth, settling and drift have a direct impact on the local molecular abundances. A key question then becomes, which molecules could be tracing the presence of embedded planets in disks and how can we disentangle this from the complex disk chemistry occurring in the background. Circumplanetary disks (CPDs) have now been detected in mm-dust emission but the optimal molecular gas tracers of these structures are still unknown \citep{2021ApJ...916L...2B, 2021arXiv210108369F}.
Forming planets are predicted to heat the disk locally, sublimating ices, causing chemical asymmetries \citep{2015ApJ...807....2C} and CPDs should in theory have a warm gas component \citep[e.g.][]{2017ApJ...842..103S,2019A&A...624A..16R}.

In this paper we present follow-up high angular resolution ALMA observations of SO in the planet-forming HD~100546 disk. 
SO has only been detected and imaged in a handful of disks \citep{2016A&A...589A..60P, 2018A&A...611A..16B, 2021A&A...651L...6B} but in younger sources ($\lesssim$1~Myr) is an important tracer commonly associated with shocks or disk winds/outflows \citep[e.g.,][]{2014Natur.507...78S, 2017MNRAS.467L..76S, 2017A&A...607L...6T,2022A&A...658A.104G}. 
In Section 2 we describe the source, HD~100546, the evidence for planets in this disk, and our ALMA observations. In Section 3 we describe the results, in Section 4 the results are discussed in context of both on-going astrochemistry and giant planet-formation in the HD~100546 disk, and in Section 5 we state our conclusions.

\begin{table*}[h!]
\begin{threeparttable}
\caption{Molecular lines and properties}
\label{table2}
\begin{tabular}{lccccccc}
\hline \hline
Transition & Frequency & $\mathrm{E_{up}}$ & \color{black} $\mathrm{A_{ul}}$ \color{black}   & Beam     & rms$^{1}$                           & Peak                               & Int. Flux$^{2}$ \\  
                 &  (GHz)    & (K)              & $\mathrm{(s^{-1})}$ &       & (mJy beam$^{-1}$) &   (mJy beam$^{-1}$) & (mJy km s$^{^{-1}}$) \\ 
\hline 

SO~~~~\ce{7_{7}} - \ce{6_6} 	& 301.286124 &  71.0 & $3.429\times10^{-4}$ & 0\farcs23 $\times$ 0\farcs19 (67$^{\circ}$) & 1.02 & 6.57 & 124 \\
SO~~~~\ce{7_{8}} - \ce{6_7}  & 304.077844  &  62.1  &  $3.609\times10^{-4}$  & 0\farcs23 $\times$ 0\farcs19 (66$^{\circ}$) & 0.92 & 6.18 & 143  \\
SO~~~~stacked  & -& - & - & 0\farcs23 $\times$ 0\farcs19 (66$^{\circ}$) & 0.69 &  11.07  & -\\
\hline
SiO~~~~7 - 6    &  303.926812   &  58.35  & $1.464\times10^{-3}$ & 0\farcs23 $\times$ 0\farcs19 (67$^{\circ}$) & 0.52  & - & $<19$ \\
SiS~~~~16 - 15  &   290.380757 & 118.47  & $4.160\times10^{-4}$  &  0\farcs23 $\times$ 0\farcs19 (70$^{\circ}$) & 0.58 & - & $<19$ \\
\hline
\end{tabular}
\begin{tablenotes}\footnotesize
\item{The values for the line frequencies, Einstein A coefficients, and upper energy levels ($E_{up}$) 
are from the Leiden Atomic and Molecular Database: \url{http://home.strw.leidenuniv.nl/~moldata/} \citep[LAMDA;][]{2005A&A...432..369S}.
$^{1}$ per 0.12 km$~\mathrm{s^{-1}}$ channel for SO and per 0.24 km$~\mathrm{s^{-1}}$ channel for SiO and SiS. All images generated with a Briggs robust parameter of 0.5. $^{2}$ extracted from a 0.6" on source elliptical aperture with the same position angle as the disk as measured from ALMA observations, e.g.,  \citet[][]{2019ApJ...871...48P}.}
\end{tablenotes}
\end{threeparttable}
\end{table*}


\section{Observations}

\subsection{The HD 100546 disk and its planets}

One system with plentiful evidence for on-going giant planet formation is the nearby (110~pc) Herbig Be star HD~100546 \citep{2018A&A...620A.128V}. 
In the outer disk a point source has been detected at a de-projected distance of $\approx60$~au that has been attributed to a giant planet called HD~100546~b \citep{2013ApJ...766L...1Q, 2015ApJ...807...64Q}. 
Additionally there is evidence for another giant planet, HD~100546~c,  orbiting within the dust cavity at $\approx10-15$~au. The main tracer of this planet is the time variable CO ro-vibrational line emission from the inner disk \citep{2009ApJ...702...85B, 2013ApJ...767..159B, 2014ApJ...791..136B, 2019ApJ...883...37B}. 
There is also a \color{black} complementary \color{black} disk feature detected scattered light data that could be the inner edge of the dust cavity or the giant planet HD~100546~c \citep[e.g.][]{2015ApJ...814L..27C, 2017RNAAS...1...40C}. 
However, there has been some debate in the literature on the detection of HD~100546~c in the scattered light data. HD~100546~c was first identified by \citet{Currie2015} and later works have investigated the robustness of the detection \citep{2017AJ....153..264F,2017RNAAS...1...40C}. The original detection of the feature in the disk is consistent with a hotspot in the disk traced via the CO P26 line (49.20~nm) at a similar time. Since 2017 the CO asymmetry is no longer detected as the putative planet is inferred to be located behind the near side of the inner rim of the disk cavity. If the scattered light feature is tracing the same feature then its expected to also disappear. More recently, HD~100546~c has indeed gone undetected with VLT/SPHERE \citep{2018A&A...619A.160S}. However, more observations and modelling work are needed to see whether or not the original feature is real and to determine if \color{black} it \color{black} is a disk feature or a planet candidate \citep{2022arXiv220505696C}.

This direct evidence for forming planets also coincides nicely with indirect evidence from ALMA observations. 
The millimeter dust in this disk has now been well studied with ALMA and is distributed in two main rings \citep{2014ApJ...791L...6W,2019ApJ...871...48P,PerezS2019,2021A&A...651A..90F}.
These rings are consistent with two giant planets embedded in the disk with modelled radial locations ranging from 10-15~au and 70-150~au
\citep{2015A&A...580A.105P, 2021A&A...651A..90F, 2021A&A...656A.150P,2022arXiv220603236C}. 
A range of molecular gas tracers have also now been detected in the HD~100546 disk. The disk is gas rich with detections of CO isotopologues \citep[\ce{^{13}CO} and \ce{C^{18}O};][]{2019MNRAS.485..739M} and the \ce{^{12}CO} emission shows non-keplerian kinematics that may be related to on-going planet formation \citep{2017A&A...607A.114W,  2019ApJ...883L..41C, 2020ApJ...889L..24P}. Low angular resolution, 1\farcs0 ALMA observations have also detected SO, \ce{H_2CO} and \ce{CH_3OH} originating from the inner $50$~au of the disk and a ring of \ce{H_2CO} (and tentatively \ce{CH_3OH}) that is co-spatial with the outer dust ring \citep{2018A&A...611A..16B, 2021NatAs...5..684B}.  In the 1\farcs0 data presented by \citet{2018A&A...611A..16B} the SO emission was spatially unresolved but there was an indication the emission was spatially asymmetric and the kinematics were clearly non-Keplerian when compared to the molecular disk traced in \ce{^{12}CO}. \citet{2018A&A...611A..16B} proposed that the SO could be tracing a disk wind, a warped disk or even \color{black} an \color{black} accretion shock onto a CPD. 
In this paper we aim to determine the origin of the SO in the HD~100546 disk with follow up high-angular resolution ALMA observations.

\subsection{ALMA Observations}

HD~100546 was observed in the ALMA Band 7 program 2019.1.00193.S (PI. A. S. Booth). The data include two configurations: 43C-2 and 43C-5. The different baselines, on source times and the spectral set up are outlined in Table A.1 in Appendix A. The data were analysed using CASA version 5.6.0 \citep{2007ASPC..376..127M}. The short baseline data were analysed in \citet{2021NatAs...5..684B} where only the \ce{H_2CO} and \ce{CH_3OH} lines were presented (i.e., not the \ce{SO}). In this work we take these self-calibrated data and combine it with the more recent pipeline calibrated long baseline observations. These combined data were then self-calibrated following same the procedure as described in \citet{2021ApJS..257....1O}. Both phase and amplitude self-calibration were applied. The continuum was then subtracted using \texttt{uvcontsub}. 

This paper focuses on the targeted molecules SO, and, SiO and SiS which are also known to be tracers of shock chemistry \citep[e.g.][see Table~1 for the specific transition information]{2017MNRAS.470L..16P}. The lines were imaged with \texttt{tCLEAN} using Briggs robust weighting, the "multi-scale" deconvolver, and a keplerian mask \footnote{\url{https://github.com/richteague/keplerian_mask}}. A range of robust parameters (+1.0, +0.5, 0.0, -0.5) were explored in the imaging. The velocity resolution of the SO spectral windows is 0.06~km~s$^{-1}$ but the data were rebinned to 0.12~km~s$^{-1}$ to improve the channel map signal-to-noise ratio. Similarly, SiO and SiS were observed at 0.12~km~s$^{-1}$ and 0.24~km~s$^{-1}$ respectively and both imaged at a spectral resolution of 0.25~km~s$^{-1}$. Following \citet{1995AJ....110.2037J} and \citet{2021ApJS..257....2C} the JvM correction was then applied to the data to account for the non-gaussianity of the dirty beam - especially important when combining data from multiple configurations. 
The typical \color{black} epsilon \color{black} value for the JvM correction was $\approx$0.50 for the Briggs robust +0.5 images. The SO lines were both well detected and the SiO and SiS were undetected. To increase the signal-to-noise in the SO image the two SO spectral window measurement sets were concatenated using CASA task \texttt{concat} and imaged together. The image properties for all of the individual and stacked lines with robust values of +0.5 are listed in Table~1. 
The 3$\sigma$ upper limits on the integrated fluxes for SiO and SiS were propagated from the rms noise in the  channel maps following the method described in \citet{2019A&A...623A.124C} and these values are listed in Table~1.

\section{Results}

\subsection{Continuum emission}

A continuum image was generated using the full bandwidth of the observations after flagging the line-containing channels. A range of robust parameters (+1.0, +0.5, 0.0, -0.5) were explored in the imaging and the resulting image properties are listed in Table B.1. The full continuum disk maps using the range of robust parameters are shown in the Appendix in Figure~B.2. These data are sensitive enough to detect the outer ring at 2\farcs0 that was predicted to be present in visibility modelling by \citet{2014ApJ...791L...6W} and first imaged by \citet{2021A&A...651A..90F}. A zoomed-in 0\farcs6$\times$0\farcs6 version of the robust = 0.5 map is shown in Figure~1. This image has a beam size of 0\farcs23$\times$0\farcs19 (63.0$^{\circ}$), a JvM epsilon of 0.56, an rms of 0.037~mJy~beam$^{-1}$ and a resulting signal-to-noise ratio of $\approx$3305.
With these high signal-to-noise data we also note the detection of a new sub-millimeter source located 9\farcs34 from HD~100546 which we discuss in Appendix C. 

\subsection{Line emission maps}

\begin{figure*}
    \includegraphics[trim=0cm 0cm 0cm 0cm, clip=true,width=0.95\hsize]{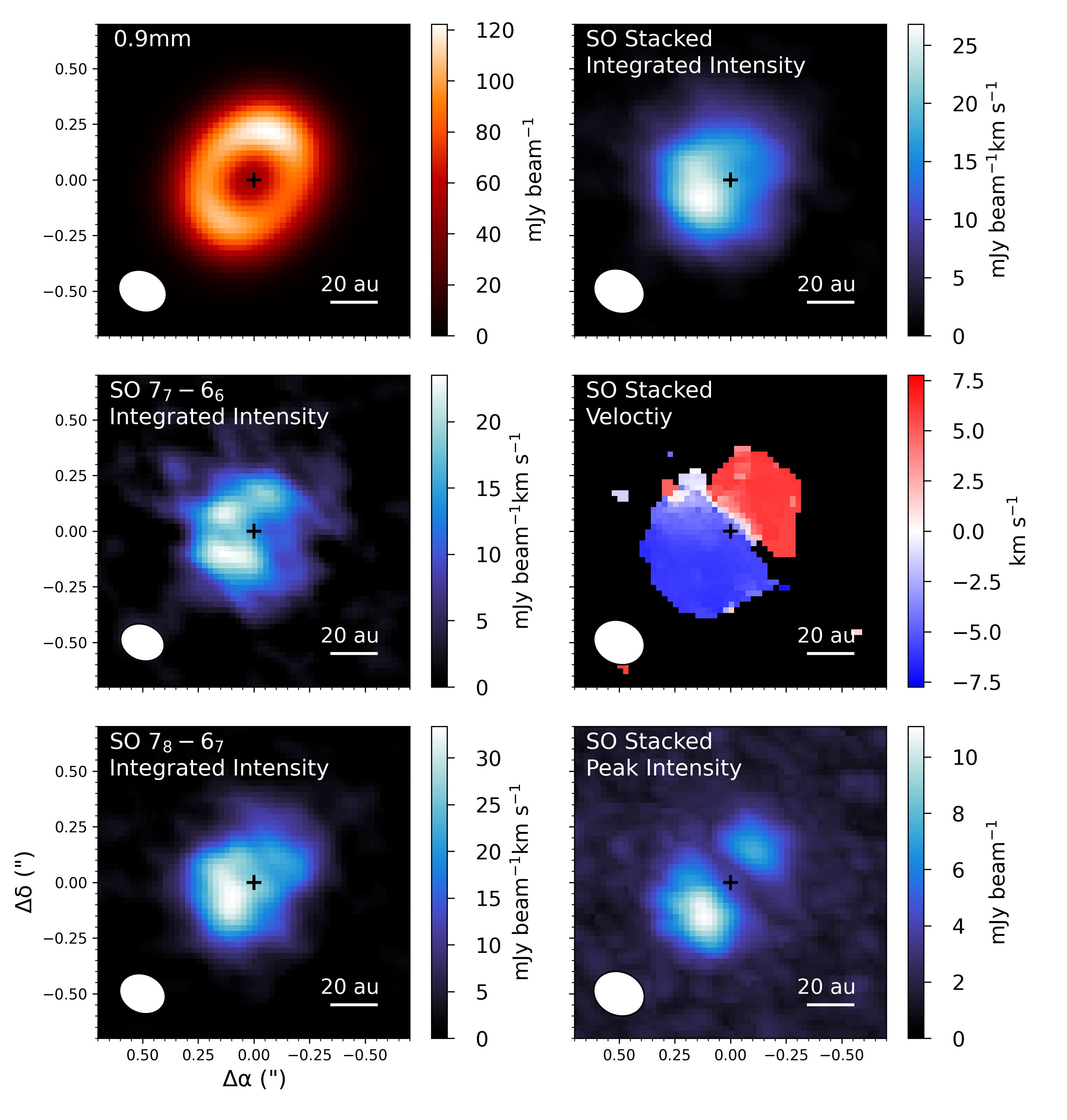}
    \caption{Emission map of the 0.9~mm continuum, Keplerian masked integrated intensity maps of the individual and stacked SO transitions, and the SO stacked intensity weighted velocity and peak emission maps. The beam size is shown in the bottom left of each panel and images were constructed using a Briggs robust parameter of +0.5.}
    \label{fig1}
\end{figure*}

The Keplerian-masked integrated intensity maps of the individual SO transitions and the stacked image with a robust parameter of +0.5 are presented in Figure~\ref{fig1} alongside the 0.9~mm continuum map. The main component of the SO emission shown in these maps has same spatial scale as the ringed dust emission. The presence of a central cavity in the SO and an azimuthal asymmetry was explored by imaging the stacked data with a range of Briggs robust parameters. The resulting Keplerian masked integrated intensity maps for a robust parameter range of [+1.0, +0.5, 0.0, -0.5] are shown in the Appendix in Figure~B.1, alongside Table~B.1, which notes the beam size and signal-to-noise ratio of each image. The asymmetry in the north west and south east of the disk is present across all of the images and the cavity becomes more pronounced with lower robust values. 
As an additional check we imaged the stacked SO before and after continuum subtraction and extracted the spectrum (as described in Section 3.3). These two spectra are shown in Figure~2 and the asymmetry persists in the image before continuum subtraction. 
Also presented in Figure~\ref{fig1} is the intensity weighted velocity map and the \color{black} peak intensity \color{black} map for the stacked image. The velocity map clearly shows that the gas is in rotation with red and blue shifted emission consistent with what has been seen in other line data of this disk \citep[e.g., \textcolor{black}{\ce{^{12}CO} $J=2-1$;}][]{2020ApJ...889L..24P}. 
The asymmetry that is seen in the integrated intensity map is much more pronounced in the \color{black} peak intensity \color{black} map with the peak emission differing in strength by approximately a factor of 2 between the north-west and south-east sides of the disk. 

\subsection{Kinematics}

\begin{figure*}
    \centering
    \includegraphics[width=0.9\hsize]{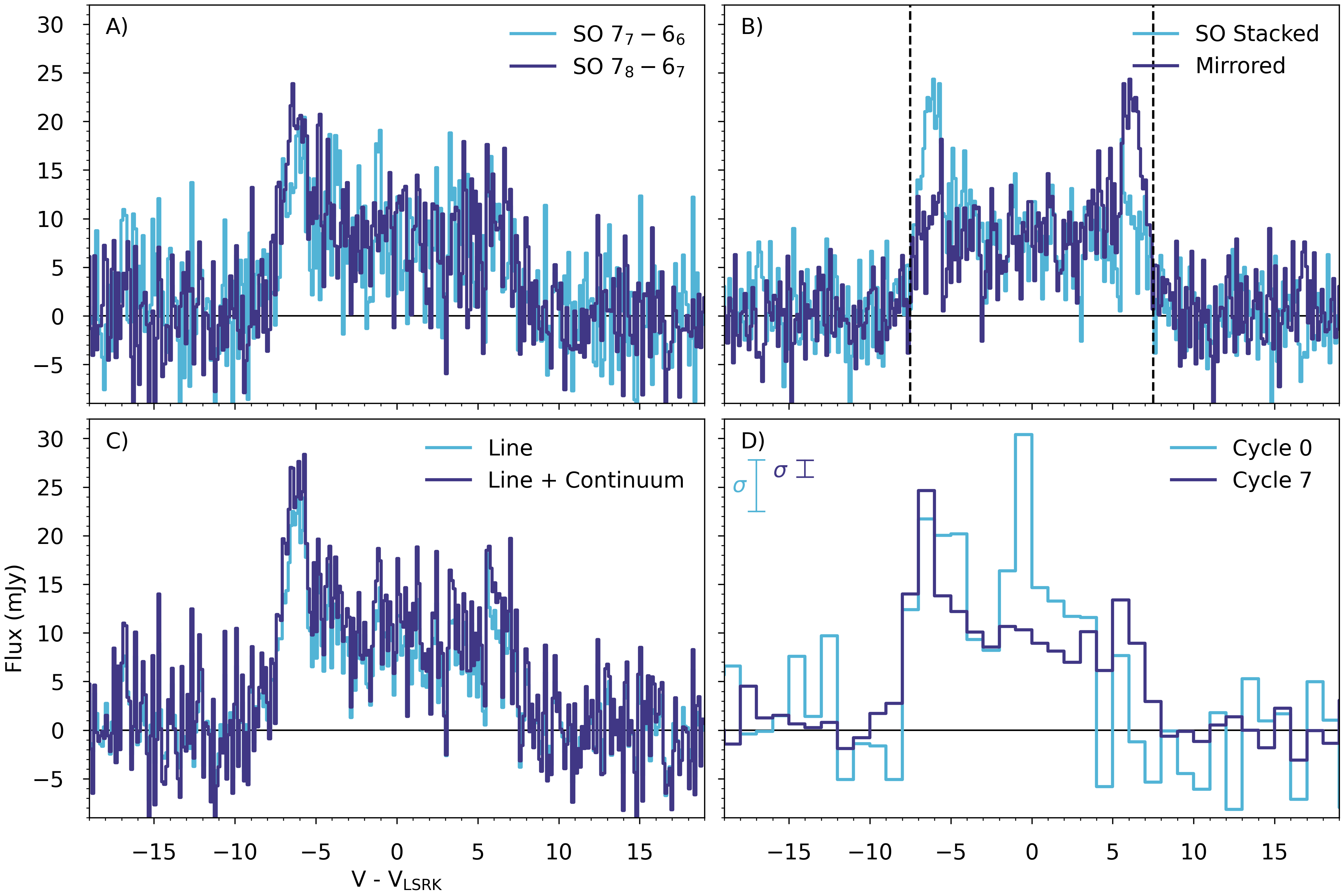}
    \caption{Spectra extracted from a 0\farcs6 ellipse. A) Individual transitions. B) Spectra of the two stacked SO lines and the same spectra mirrored. C) Spectra of the two stacked SO lines with continuum subtraction and without continuum subtraction in the UV domain (after subtracting average value of line free continuum emission channels). D) Spectra of the two stacked SO lines from the ALMA Cycle 0 observations presented in \citet{2018A&A...611A..16B} and the Cycle 7 observations presented in this work. The Cycle 7 data have been imaged at 1.0~km~s$^{-1}$ in order to make a direct comparison to the Cycle 0 data. The noise calculated from the line free channels in each respective spectrum is shown by a bar in the top left corner of the plot. }
    \label{fig2}
\end{figure*}

The spectra for each transition are extracted from an ellipse with the same position and inclination angle as the disk and a semi-major axis of $0\farcs6$, and are shown in Figure~\ref{fig2}. Both line profiles have a very similar, asymmetric shape. 
We calculate the integrated lines fluxes (listed in Table~1) for the two SO transitions by integrating the line profiles between $\pm7.5$~km~s$^{-1}$. The line profile from the stacked image is shown in Figure~\ref{fig2} alongside a mirrored version. This shows that the line is symmetric in width about the source velocity (5.7~km~s$^{-1}$, \citealt{2017A&A...607A.114W}). This was not the case in the Cycle 0 data presented in \citet{2018A&A...611A..16B} (see panel D in Figure~2 for a comparison, this will be discussed further in Section~4.3). The emission drops off sharply for both the blue and red shifted sides of the disk at an average of $7.5\pm0.12$~km~s$^{-1}$ with respect to the source velocity. Assuming Keplerian rotation, an inclination angle of 32$^{\circ}$ (as derived by \citet{2019ApJ...871...48P} for the inner 2\farcs0 of the disk), and stellar dynamical mass of 2.2~M$_{\odot}$ results in a inner ring radius of $\approx$9.8~$\pm0.3$au. If the inclination in this region of the disk is higher (40$^{\circ}$; \citealt{2021arXiv211200123B}) then this results in a larger inner radius of $\approx$14.5~$\pm0.3~$au. The errors here are propagated from one velocity channel. 

\subsection{Radial emission profile}

Further information on the SO emitting region(s) can be gained from the spectra. Using the tool GoFish\footnote{\url{https://github.com/richteague/gofish}} \citep{GoFish} we shift and stack the spectra from each pixel in the channel maps. The spectra are extracted from annuli with width 0\farcs05.   
This analysis revealed a clear second ring of SO in the outer disk and the resulting radial emission profile for the stacked image are shown in Figure~3. 
The errors in this profile are calculated per radial bin from the rms in the shifted spectra. 
This outer ring is also tentatively present in the azimuthally averaged radial profiles from the intensity maps (shown in the Appendix in Figure B.3) and the spectral stacking confirms this feature. 
In Figure~3 we also compare these line radial profiles to that of our 0.9~mm continuum data imaged at the same weighting. The second SO ring is peaking slightly beyond the peak of the second dust ring at $\approx$200~au. These profiles are extracted assuming a flat emitting surface.
\textcolor{black}{We do not have high enough spatial resolution to constrain an emission surface for the SO emission; however, as discussed in \citet{2021ApJS..257....3L}, for more extreme emission heights of $z/r \approx 0.1-0.5$ the shape and structure of the radial emission profiles are not significantly altered.}
\color{black}


\subsection{Column density of SO}

The radial emission profiles can be converted to a radial column density of SO under a few assumptions. Following a similar analysis as, e.g., \citet{2018ApJ...859..131L} and  \citet{2021ApJ...906..111T}, we assume that the lines are optically thin, in local-thermodynamic equilibrium and, since we do not have enough transitions detected over a range of upper energy levels in order to perform a meaningful rotational diagram analysis, we fix the rotational temperature ($T_{\mathrm{rot}}$). For the inner disk ($<$100~au) $T_{\mathrm{rot}}$ is chosen as 75, 100, and 200~K and in the outer disk ($>$100~au) 30, 40 and 50~K are chosen. These are reasonable given the values of the dust and gas temperatures in the HD~100546 physical model from \citet{2016A&A...592A..83K}. The resulting radial column density profile is the average of the two detected transitions and is shown in Figure~3, where the errors are propagated from the errors in the stacked radial emission profile. For all values of $T_{\mathrm{rot}}$ the line is optically thin across the entirety of the disk. If the width of the emitting area of the SO in the inner ring were much narrower than we observe then the optical depth and therefore column density we derive may be underestimated.  

The azimuthal asymmetry is particularly interesting to investigate to see how this variation \color{black} affects \color{black} the column density in the inner disk. Azimuthal profiles are extracted from the integrated intensity maps at 20~au (the radius at which the SO peaks) and converted to column density. The azimuthal variation in integrated flux and column density of a factor of 2 is seen in Figure~4. This could either be due to a chemical asymmetry, i.e., there are more SO molecules on the southern side of the disk than the northern, or a radiative transfer effect due to a temperature asymmetry. If the column density of SO is constant around the inner disk then a temperature variation of $\approx \times 2$ is required between the north and south of the disk, assuming that the emission is indeed optically thin. 

We also estimate the average SO column density in the inner disk and the upper-limits on the SiO an SiS column densities. This was done using the integrated fluxes as listed in Table 1 and an assumed emitting area of the 0\farcs6 ellipse ($1.8~\times~10^{-11}$ steradians) from which the spectra were extracted. We note that these average values will underestimate the column density when compared to the radially/azimuthally resolved analysis. However, these are reasonable values to calculate in order to put upper limits on the SiO/SO and SiS/SO ratios and better place the SO detection in context with the range of non-detections in the literature. The results of these calculations are listed in Table 2 where the ratios of the column densities are SiO/SO~$<$~1.4\% and SiS/SO~$<$~5\%.

\section{Discussion}

\subsection{Morphology of the SO emission}

The SO emission detected in the HD~100546 disk is constrained to two rings. This double ring structure closely follows that of the millimeter dust disk. From the kinematics, the SO emission is detected down to $\approx 10-15$~au and is therefore tracing gas within the millimeter dust cavity   \citep[e.g.][]{2019ApJ...871...48P}. There is a clear brightness asymmetry in the inner ring of SO of a factor of 2. 
The outer SO ring is located from 160 to 240~au, peaking just beyond the second dust ring, and is a factor of 100 times lower in peak flux than the inner ring. In comparison, the contrast ratio between the inner and outer dust rings is $\approx 1000$. There is also a shoulder of SO emission outside of the inner dust ring out to $\approx 90$~au.  In the following sections we discuss the chemical and/or physical processes that could lead to the distribution of SO that we observe in the HD~100546 disk.

\begin{figure*}[h!]
    \centering
    \includegraphics[width=0.9\hsize]{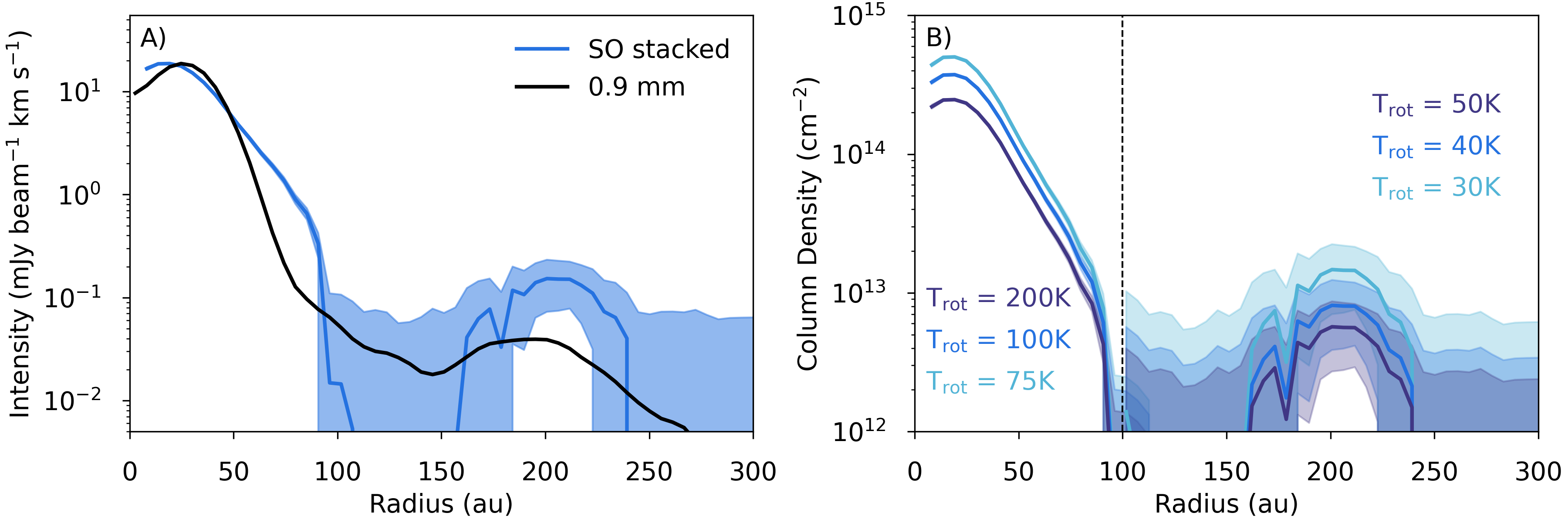}
    \caption{A) Azimuthally averaged intensity profiles for the stacked SO emission and 0.9~mm continuum emission. B) Radial SO column density calculated at a range of rotational temperatures.}
    \label{fig3}
\end{figure*}

\begin{figure*}[h!]
    \centering
    \includegraphics[width=0.9\hsize]{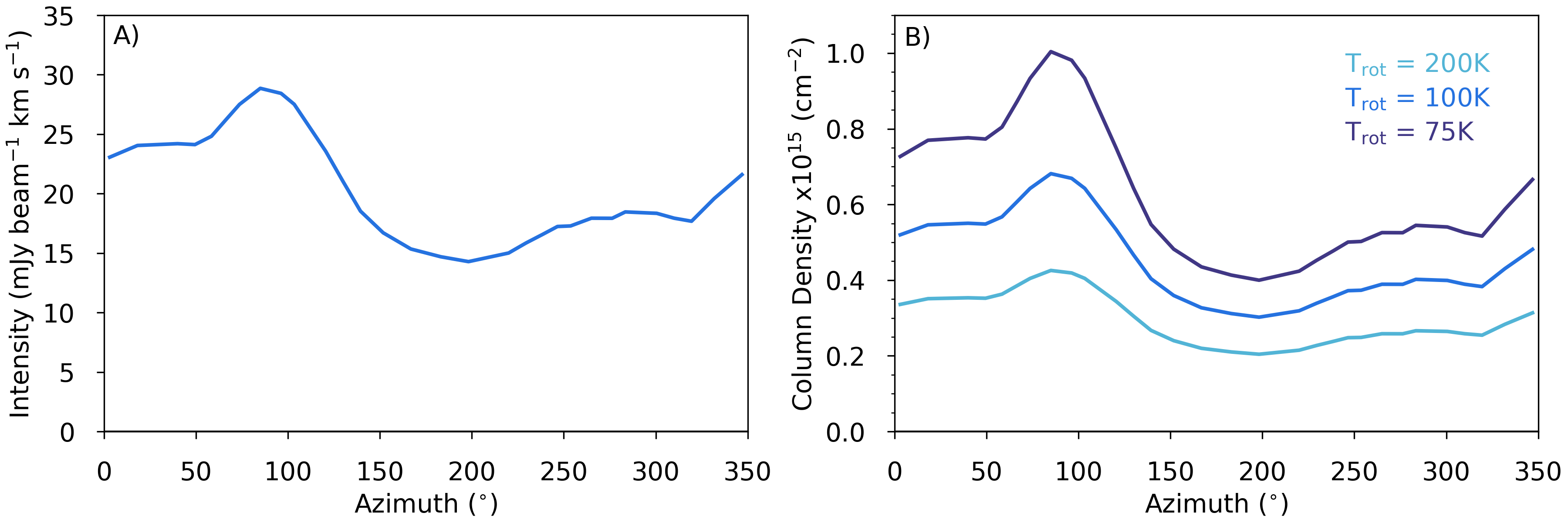}
    \caption{A) Azimuthal intensity profile for the stacked SO emission at ~20~au. B) Azimuthal column density calculated at a range of rotational temperatures at 20~au.}
    \label{fig 4}
\end{figure*}

\begin{table}[]
    \centering
    \caption{Average column densities for the inner 0\farcs6 of the HD~100546 disk for the detected and non-detected molecules.}
    \begin{tabular}{cccc}
    \hline \hline
Molecule  & $\mathrm{T_{ex}}$ (K) & $\mathrm{N_{T}}$ (cm$^{\mathrm{^{-2}}}$) \\ \hline
    \ce{SO} & 75-200K &  $4.0-6.4$ $~\times~10^{13}$ \\
    \ce{SiO} & 75-200K &  < $4.9-7.9$ $~\times~10^{11}$  \\
    \ce{SiS} &75-200K &  < $2.0-3.2$ $~\times~10^{12}$ \\
 \hline
     \ce{SiO/SO} & &  < 1.4\% \\
     \ce{SiS/SO} &  &  < 5.0\% \\
   \hline \\
        \end{tabular}
\end{table}

\subsection{Chemical origin of the SO}

SO has been proposed as a tracer of a variety of physical and chemical processes across star, disk, and planet formation including (accretion) shocks, proto-stellar outflows, MHD driven disk winds and the sublimation of ices around hot cores \citep{2014Natur.507...78S, 2017MNRAS.467L..76S, 2017A&A...607L...6T, 2018ApJ...863...94L, 2020A&A...637A..63T, 2021A&A...653A.159V, 2021A&A...655A..65T}. However, with that being said, detections of SO in the more evolved Class II planet-forming disks are scarce. HD~100546 is one of only three disks where SO has been imaged, the others being AB~Aur and IRS~48 \citep{2016A&A...589A..60P, 2021A&A...651L...6B}. As well as the general problem in astrochemistry of sulphur depletion in dense gas \citep[e.g.][]{1994A&A...289..579T,2019A&A...624A.108L} the lack of SO in many disks has been linked to an overall observed depletion of volatile oxygen relative to carbon  \citep[e.g.][]{2017A&A...599A.113M}. The SO to CS ratio has therefore been explored as a tracer of the gas-phase C/O ratio in disks \citep{2018A&A...617A..28S, 2019ApJ...876...72L, 2020A&A...638A.110F, 2021ApJS..257...12L}. For example, \citet{2018A&A...617A..28S} show that the relative column densities of SO and CS decrease and increase by a factor of $\approx100$ respectively when changing the elemental C/O ratio from 0.45 to 1.2. This means that at an elevated C/O ratio CS traces the bulk of the gas phase volatile sulphur in disks.
This is consistent with observations of CS in disks where \ce{C_2H}, another tracer of an elevated C/O environment, is detected \citep[e.g.][]{2021arXiv210108369F}.
However, the inferred S/\ce{H} ratios from modelling CS emission from disks are low, $\approx~8~\times~10^{-8}$, indicating that a significant S reservoir is still locked in the ices or a more refractory component \citep[e.g.][]{2021ApJS..257...12L}.  
There are some caveats in the chemical models of sulphur in disks, namely, the initial partitioning of S in the ices, the completeness of the sulphur grain surface reaction networks, and photo-desorption yields that remain uncertain. 
All of these processes and assumptions are important if the observed gas-phase molecules have an ice origin. 
The level of oxygen (and carbon) depletion in HD~100546 has been modelled by \citet{Kama2016twhya} who find no evidence that oxygen is significantly more depleted than carbon throughout this warm disk unlike in other, cooler disks, e.g., TW Hya \citep{Kama2016twhya}. This leads to the expectation that there is some level of gas-phase SO present through the entire disk. 

From the high column density, and given that the location of the SO emission is from the inner region of the HD~100546 disk, it is most likely that the SO originates from thermal desorption of S-rich ices. This is consistent with the evidence for warm $100-300$~K dust and compact methanol emission from this same region of the disk \citep{2011A&A...531A..93M, 2021NatAs...5..684B}. This release of S-ices at the edge of an inner dust cavity has been proposed by \citet{2019ApJ...885..114K}, although the exact form of the S-ices (e.g., \ce{H_2S}, OCS, or \ce{SO_2}), is still unclear. SO can form efficiently in the gas phase via \color{black} barrierless neutral-neutral reactions\footnote{\color{black}KIDA: KInetic Database for Astrochemistry \color{black}}, S + OH  or O + SH  if \ce{H_2O} and \ce{H_2S} ices are desorbed from the grains and subsequently photo-dissociated \cite[see the discussion in][]{2018A&A...617A..28S}:  
\begin{align}
\ce{H2O + h\nu} & \longrightarrow \ce{OH + H} \nonumber \\
                & \longrightarrow \ce{O + H2} \nonumber \\
\ce{H2S + h\nu} & \longrightarrow \ce{HS + H} \nonumber \\
                & \longrightarrow \ce{S + H2} \nonumber
\end{align}
followed by
\begin{align}
\ce{S + OH}     & \longrightarrow \ce{SO + H} \nonumber \\
\ce{O + HS}     & \longrightarrow \ce{SO + H}. \nonumber 
\end{align}
\color{black}
OH has indeed been detected in the HD~100546 disk with \textit{Herschel}/PACS and originates from warm (200~K) gas \citep{2013A&A...559A..77F}.

This ice sublimation scenario would predict an SO distribution that follows the large dust grain distribution and therefore should be broadly azimuthally symmetric; potential reasons for the observed asymmetry will be discussed in Sections 4.3 and 4.4. 
Overall, the SO (and \ce{CH_3OH}; \citealt{2021NatAs...5..684B}) in the inner HD~100546 disk is similar to what has been observed in IRS~48 \citep{2021A&A...651L...5V,2021A&A...651L...6B}.  In both of these warm Herbig transition disks ice sublimation rather than gas-phase chemistry appears to be responsible for most of the observable molecular column density. This is likely due to the close proximity of the dust (and ice) traps to the central stars. Future observations of other molecules will help to put these disks in better context other well characterised Herbig disks, e.g., HD~163296 and MWC~480 \citep{2021ApJS..257....1O}. 

\color{black} Due to the much colder temperatures in the outer disk, the ring of SO that is co-spatial/beyond the dust ring at 200 au will have a different chemical origin. \color{black} This is likely UV and/or cosmic ray triggered desorption from icy dust grains. 
\color{black}
This would be similar to the explanation for the origin of \ce{H_2CO} emission in several disks, which shows an increase in column density at the edge of the millimeter dust disk \citep{2015ApJ...809L..25L, 2017A&A...605A..21C, 2018ApJ...863..106K, 2020ApJ...890..142P, 2021ApJS..257....6G}.
Additionally, in a few sources, CS or SO is detected in a ring, approximately co-spatial with \ce{H_2CO} emission, beyond the mm-dust disk \citep{2018ApJ...863..106K, 2019ApJ...876...72L, 2020A&A...644A.119P, 2020A&A...642A..32R}. \color{black}
There are also rings of \ce{H_2CO} and \ce{CH_3OH} in this region of the HD~100546 disk which would be consistent with this scenario \citep{2021NatAs...5..684B}.
Clear rings of SO and \ce{H_2CO} have also been observed in AB~Aur just beyond the dust trap \citep{2020A&A...642A..32R}.  Although SO is detected in AB~Aur the models explored in \citet{2020A&A...642A..32R} show that the AB~Aur data are most consistent with an elevated C/O = 1, a \color{black} depleted sulphur abundance of $8\times10^{-8}$ and an SO \color{black} abundance of $4\times10^{-10}$. The environment in the HD~100546 outer disk could be similar to this but further models are needed to derive the gas-phase C/O and S/H across this disk. 

In HD~100546, the depletion of SO in the dust gap, and in the outer disk where CO is still abundant, may be an indication of an elevated C/O ratio in the disk gas. This could be explained by volatile transport and an enhancement of S and O-rich ices in the mm-dust rings with peaks of SO emission just outside the rings and depletion in the regions without mm-dust. 
Additionally/alternatively, if SO is present in the gas-phase throughout the entire gas disk there could be a depletion of total gas column density in the outer mm-dust gap potentially reducing the SO column below our detection limit; however,  no clear CO gaps have been detected \citep[e.g][]{2019ApJ...871...48P}. 
It is interesting to note that \ce{H_2O} ice and emission from cold \ce{H_2O} gas has been detected within the gap region from $\approx$40-150~au
(\citealt{2016ApJ...821....2H}, \citealt{2021A&A...648A..24V},\citealt{2022arXiv220710744P}).
Dedicated models of chemistry in planet-carved dust gaps show that it is important to also consider chemical processing in dust gaps which will also \color{black} affect \color{black} the relative abundances of different gas tracers \citep{2021ApJS..257....8A}.
Further disk specific chemical models and observations of complementary molecules are needed to better constrain the chemical conditions in the HD~100546 disk. 

\subsection{Origin of the SO asymmetry}

The bulk of the SO emission in the HD~100546 disk is tracing gas just within the millimeter dust cavity. The asymmetry in the detected emission ring could be due to an azimuthal variation in the SO abundance and/or disk temperature. This is distinct to the CS asymmetries presented in \citet{2021ApJS..257...12L} for the five disks in the sample from the ALMA large program "Molecules with ALMA at Planet-Forming Scales" (AS209, GM~Aur, IM~Lup, HD~163296, MWC~480). In these other disks the asymmetry is across the near and far sides of the disks, so could just be due to our viewing angles of the systems; however, in HD~100546 the asymmetry is not across this axis. In this section we discuss potential origins of the SO asymmetry in the HD~100546 disk. 

In \citet{2018A&A...611A..16B} the SO emission was attributed to a disk wind due to excess blue-shifted emission in the line profile (see Figure~2). In those Cycle 0 data the emission was spatially unresolved and the kinematics appeared non-keplerian. The stacked spectra from \citet{2018A&A...611A..16B} are shown in Figure~2 where the emission peaks at the source velocity and the line profile has an excess blue-shifted component. This interpretation of the data was very plausible since it has been shown that HD~100546 drives a jet \citep{2020A&A...638L...3S} and SO has also been shown to be a tracer of MHD driven disk winds \citep{2020A&A...640A..82T}. The stacked spectra from the new Cycle 7 data presented in this work are compared to the Cycle 0 data in Figure~2. In this Figure, the Cycle 7 data have been imaged at 1.0~km~s$^{-1}$ in order to make a direct comparisons to the Cycle 0 data. The two line profiles are noticeably different. 
The Cycle 0 data were taken in November 2012 and the Cycle 7 data in June 2021, i.e., over a $\approx$8.5 year baseline. 
\textcolor{black}{Figure B4 in the appendix shows the residual spectra of the Cycle 7 spectra subtracted from the Cycle 0 spectra. The emission in the central channel, where $\mathrm{V=V_{LSRK}}$ remains above the $\pm$3~$\sigma$ level. Where $\sigma$ is the rms of each respective spectra added in quadrature}.
The overall line profile in the Cycle 7 data is symmetric in velocity about the source velocity compared to the Cycle 0 spectra. 
The Cycle 7 data therefore do not strongly support the disk wind hypothesis. Whilst we 
still see a brightness asymmetry we do not see an additional component of emission that could be attributed to a disk wind. 

A potential origin for the SO brightness asymmetry is an azimuthal temperature variation which could be caused by a warp. An extreme warp in the inner 100~au of the disk was suggested by \citet{2017A&A...607A.114W} from ALMA observations of \ce{^{12}CO} and previously a warp was suggested by \citet{2010A&A...519A.110P} as an interpretation for asymmetric line observations from APEX. A similarly asymmetric line profile has been traced in APEX CO $J=6-5$ data also \citep{2016A&A...588A.108K}. However, a warped disk has not been directly confirmed in higher resolution ALMA line observations \citep[e.g., ][]{2019ApJ...871...48P, PerezS2019,2022arXiv220603236C}. 
SO has been shown to be a potential chemical tracer of a warped disk by \citet{2021MNRAS.505.4821Y} where the variation SO abundance is primarily driven by the changes in X-ray illumination around the disk. In these generic models the warp is caused by a 6.5 $\mathrm{M_{Jup}}$ planet embedded in a disk at 5~au on a 12$^{\circ}$ misaligned orbit \citep{2019MNRAS.484.4951N}.
On sub-au scales the inner disk has been traced by \citet{2021arXiv211200123B} via VLTI/GRAVITY observations but the position and inclination angles vary only by $\approx$4$^{\circ}$ compared to the outer disk traced in CO. It is unclear if a warp of this degree could create the temperature/chemical gradient we observe. In addition, current warp chemical models do not have large mm-dust cavities like that observed in the HD~100546 system \citep[e.g.,][]{2021MNRAS.505.4821Y}. This different disk structure would be expected to alter the signature of the warp in the line emission.


SO is also a known tracer of shocks, and in particular has been proposed as a tracer of accretion shocks at the disk envelope interface in younger class 0/I sources \citep{2014Natur.507...78S, 2015ApJ...799..141A, 2017MNRAS.467L..76S, 2019A&A...626A..71A, 2021A&A...653A.159V, 2021arXiv211013820G}. SO is therefore a potential tracer of accretion shocks into a CPD. So far, there are only upper limits on the CPD dust mass in the HD~100546 cavity \citep{2019ApJ...871...48P}. 
In our data we also have a non-detection of another shock tracer SiO with an average SiO/SO ratio of $<$1.4\%. This is an indication that if \color{black} SO is \color{black} enhanced in the inner disk due to a shock, then the strength of the shock is not sufficient for the sputtering of the grain cores, and is limited to the removal of ices. In this case, a weak shock is almost chemically indistinct from thermal ice sublimation \color{black} at warm temperatures ($\gtrsim$ 100~K) because in both scenarios, the full ice mantle is removed from the dust grains without chemical alteration\color{black}. Other tracers of gas within the cavity have shown asymmetries. The CO ro-vibrational lines observed with CRIRES have been shown to trace the warm gas from the inner rim of the outer disk \citep{2014A&A...561A.102H}. The Doppler shift of the CO v=1-0 P(26) line has been shown to be time variable tracing a hot spot of CO gas that is now behind the near side of the cavity wall  \citep{2009ApJ...702...85B, 2013ApJ...767..159B, 2014ApJ...791..136B, 2019ApJ...883...37B}. The interpretation of this emission is hot gas associated with a CPD of a super-Jovian planet. There is also an asymmetry in the OH emission observed with CRIRES that matches the blue-shifted asymmetry in the SO emission, but this asymmetry in the CRIRES data can be explained as an instrumental effect when the slit width, which was smaller than the diameter of the cavity is misaligned \citep{2015ApJ...800...23F}. This is not an issue for the CO observations which were taken with a slit size larger than the cavity diameter \citep{2019ApJ...883...37B}.

\begin{figure*}
    \centering
    \includegraphics[width=0.8\hsize]{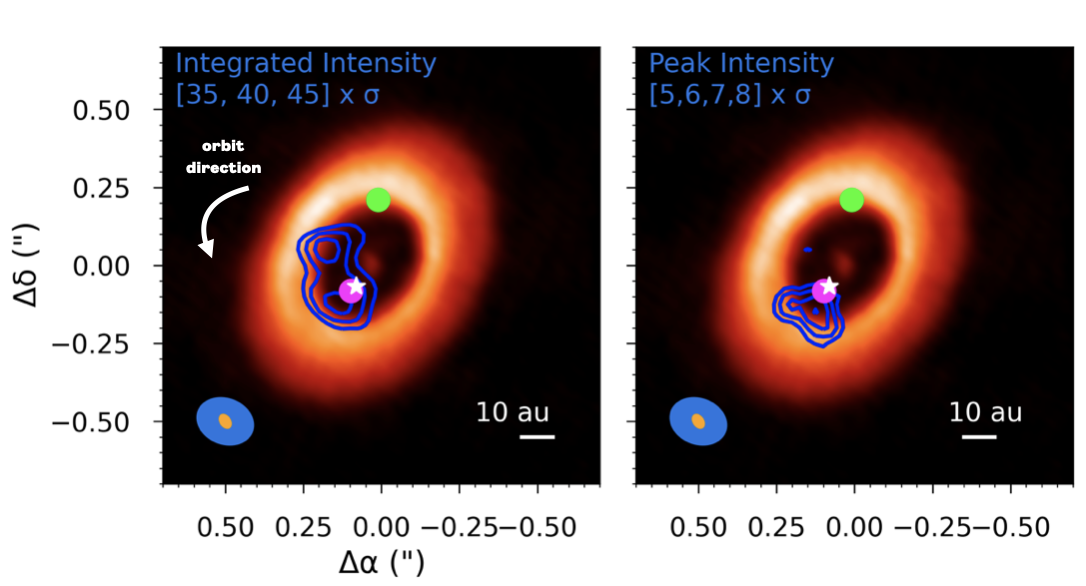}
    \caption{A composite image showing the 0.9~mm dust emission from \citet{2019ApJ...871...48P}, the [30,35,40]$\times \sigma$ (where $\sigma$ is the rms noise of an \color{black} integrated intensity map \color{black} generated without a Keplerian mask over the same number of channels) contours of the stacked SO Keplerian masked integrated intensity map (left) and the [5,6,7,8]$\times \sigma$ (where $\sigma$ is the channel map noise) contours of the \color{black} peak intensity \color{black} map (right) imaged with Briggs robust parameter of 0.0. The location of putative planet HD~100546~c (purple) \citep{2015ApJ...814L..27C} and putative planet from \citep{2019ApJ...883L..41C} (green) are highlighted with circles. The excess CO signal from \citet{2019ApJ...883...37B} in 2017 is shown by the white star symbol. Beams for the continuum observations (orange) and the line data (blue) are shown in the bottom left corner. The arrow marks the orbital direction of the disk material and planet(s). }
    \label{fig:planet}
\end{figure*}

\subsection{Connections to ongoing giant planet formation in the disk}

There are multiple complementary observations of the HD~100546 disk that suggest the presence of a giant planet within the millimeter dust cavity with an inferred orbital radius of HD~100546c~c of  $\approx 10-15$~au. Here we discuss how the asymmetry in the SO emission may be related to this putative forming planet. Figure~\ref{fig:planet} shows the high resolution 0.9~mm continuum data from \citet{2019ApJ...871...48P} with a contour overlay of the stacked SO integrated intensity and \color{black} peak intensity \color{black} maps imaged with a Briggs robust parameter of 0.0. Also highlighted are the locations of three features in the disk that are potentially linked to forming giant planets. The first is a point source identified via observations with the Gemini Planet Imager that has been attributed to either the inner edge of the dust cavity wall or forming planet HD~100546~c \citep{2015ApJ...814L..27C}.
Additionally, the CO ro-vibrational observations are
compatible with an orbiting companion within the cavity where the excess CO emission is proposed to be tracing gas in a CPD \citep{2009ApJ...702...85B, 2013ApJ...767..159B, 2014ApJ...791..136B, 2019ApJ...883...37B}. The $\approx 15$~years baseline of CO observations have allowed for an inferred orbital radius of 10.5 to 12.3~au. This is approximately the same distance from the star as the region identified as HD~100546~c. 
The third feature is a potential planet detected via the Doppler flip in the \ce{^{12}CO} kinematics as observed by ALMA.
\citealt{2019ApJ...883L..41C} and \citealt{2022arXiv220603236C} attribute this doppler flip to a forming planet/planetary wake originating from an annular groove inside the main continuum ring at $\approx 25$~au. 

The peak of the SO emission is co-spatial with the location of HD~100546~c according to the scattered light observations of \citet{2015ApJ...814L..27C} and the excess CO signal in 2017 from \citet{2019ApJ...883...37B}. Forming planets in disks have been shown in models to heat the local disk environment due to their inherent accretion luminosity and this will \color{black} affect \color{black} the observable disk chemistry \citep{2015ApJ...807....2C}. In the chemical models from \citet{2015ApJ...807....2C} SO is not specifically explored as a tracer of this ice sublimation but \ce{H_2S} and \ce{CS} do show enhancements in abundance and line strength in the same azimuthal region as the planet. The optimal tracers of this process will strongly depend on where the planet is located in the disk and the temperature, e.g., if it is warm enough to liberate \ce{H_2O} ice. However, overall an enhancement in SO in the disk due to HD~100546~c is a plausible explanation for our observations. The SO observations from Cycle 0 and 7 are taken 8.5 years apart this is $\approx 1/5$th of an orbit for a planet at 10~au. Therefore, some variation in the SO emission detected at these two epochs, due to a planet in the cavity, is feasible. Indeed we see that the spectra presented in panel D of Figure 2 are different. The movement of the peak in the spectra from $\approx 0$~km~s$^{-1}$ to $\approx -7$~km~s$^{-1}$ is consistent with the anti-clockwise rotation of a hot spot in the disk from approximately the minor axis to the south of the disk. \citet{2018A&A...611A..16B} showed that the SO emission was asymmetric peaking on the east side of the disk minor axis, and this was also traced by a brightness asymmetry in the optically thick \ce{^{12}CO} J=3-2 emission \citep{2017A&A...607A.114W}. With the SO peaking now the south of the disk in the Cycle 7 observations this is generally consistent with rotation of a hot-spot of molecular gas in the cavity/at the cavity edge; however, this signature seems to be trailing that of the CO ro-vibrational lines. SO could therefore be tracing the chemistry in the post shock regime, so behind the orbit of the planet. The CPD models of \citet{2017ApJ...842..103S} for 10~M$_{\mathrm{Jup}}$ planets, like that expected for HD~100546~c,  show very high temperatures within 0.1 times the hill sphere ($> 1000$~K) allowing dust to sublimate and likely destroying most molecules. However, in the outer regions of the CPD, the temperature can drop to a few 100~K, and with densities of $\approx 10^{14}$ cm$^{-3}$, this could be a favourable environment for \color{black} a rich, \color{black} high-temperature gas-phase chemistry. 
The SO may not directly trace accretion shocks from the HD~100546~c CPD, especially as SiO is not detected in these data, but rather the impact of HD~100546~c on the surrounding disk, and in particular, the exposed cavity wall. On this theme, it is interesting to note that the presence of a giant planet in a disk can excite the orbits of planetesimals embedded in the gas disk resulting in bow shock heating that can evaporate ices \citep{2019ApJ...871..110N}.

\color{black}

\section{Conclusions}

This paper presents ALMA observations of sulphur monoxide emission from the HD~100546 disk at $\approx 20$~au resolution showing that the dust and this gas tracer are closely linked. The SO is detected in two main rings which follow the ringed distribution of the mm-dust disk. These two rings of molecular emission are linked to ices in the disk but have different chemical origins: thermal and non-thermal ice sublimation. The inner ring gas is detected down to $\approx 10-15$~au which is within the mm-dust cavity. The mechanism of thermal sublimation of S-rich ices (e.g., \ce{H_2S}) in this region of the disk is consistent with the detection of gas phase \ce{CH_3OH} which requires an ice origin \citep{2021NatAs...5..684B}. In the outer disk the SO is co-spatial with rings of \ce{H_2CO} and \ce{CH_3OH} which all likely originate from non-thermal desorption \citep{2021NatAs...5..684B}. However, disk specific gas-grain chemical models are needed to confirm this and additional models are required to determine the C/O and S/H ratio of the gas across the disk.  The asymmetry in the SO emission from the inner disk might be directly or indirectly tracing the impact of the forming giant planet HD~100546~c on the local disk chemistry. 
Observations of other gas tracers at sub-mm wavelengths, e.g., higher $J$ CO lines, and observations of the cavity gas with complementary telescopes, e.g., VLT/CRIRES+ will help to determine if the SO is indeed a molecular tracer of giant planet formation. Furthermore, chemical and radiative transfer models to investigate the survival and observability of molecules in CPDs and the disk region within the sphere of influence of a forming giant planet are also needed to test our hypothesis put forward for a planet origin of the SO asymmetry in the HD~100546 disk. 

\begin{acknowledgements}

Authors thank Prof. Michiel Hogerheijde, Dr. Miguel Vioque, Dr. Alison Young and Dr. Sarah K. Leslie for useful discussions. 
Astrochemistry in Leiden is supported by the Netherlands Research
School for Astronomy (NOVA), by funding from the European Research
Council (ERC) under the European Union’s Horizon 2020 research and
innovation programme (grant agreement No. 101019751 MOLDISK).
ALMA is a partnership of ESO (representing its member states), NSF (USA) and NINS (Japan), together with NRC (Canada) and NSC and ASIAA (Taiwan) and KASI
(Republic of Korea), in cooperation with the Republic of Chile. The Joint ALMA
Observatory is operated by ESO, AUI/ NRAO and NAOJ. 
CW acknowledges support from the University of Leeds, STFC, and UKRI (grant numbers ST/T000287/1, MR/T040726/1)
This paper makes use
of the following ALMA data: 2011.0.00863.S, 2015.1.00806.S, 2019.1.00193.S.

\end{acknowledgements}

\bibliographystyle{aa} 
\bibliography{aanda} 

\onecolumn
\newpage
\begin{appendix}

\section{Observing set up}

\begin{table}[h!]
    \caption{Observing set up for ALMA Band 7 program 2019.1.00193.S (PI: A. S. Booth)}
    \centering
    \begin{tabular}{c c c c}
    \hline\hline
    \multicolumn{4}{c}{Short baseline data (43C-2)} \\
    \multicolumn{4}{c}{Date Observed = [10/12/2019]} \\
    \multicolumn{4}{c}{On Source time = [2068.9] seconds} \\
    \multicolumn{4}{c}{No. antenna = [43]} \\
    \multicolumn{4}{c}{Baselines = [15-312] m} \\
    \hline 
    \multicolumn{4}{c}{Long baseline data (43C-5)} \\
    \multicolumn{4}{c}{Dates Observed = [23/05/2021, 25/06/2021, 26/06/2021, 29/06/2021]} \\
    \multicolumn{4}{c}{On Source times =  [4764.48, 4654.08, 4522.27, 4665.36] seconds} \\
    \multicolumn{4}{c}{No. antenna = [39, 38, 39, 42]} \\
    \multicolumn{4}{c}{Baselines = [22-2020, 23-2010, 15-1994, 15-2517]~m} \\
    \hline 
SPW-id & Central Frequency (GHz)	& Native Resolution (km~s$^{-1}$)& Bandwidth (MHz) \\ \hline \hline 
25 &	301.2848802&	0.061&	117.2  \\ 
27 &	303.9255271	&0.060&	58.6 \\ 
29&	304.0765586	&0.060&	58.6 \\ 
31&	290.2475581	&0.126&	58.6 \\ 
33&	290.2629000&	0.126&	58.6 \\ 
35&	290.3061519&	0.126&	58.6  \\ 
37&	290.3795161&	0.126&	58.6  \\ 
39&	291.2365945&	0.126&	58.6  \\ 
41&	291.3792947&	0.126&	58.6  \\ 
43&	291.3830789&	0.126&	58.6  \\ 
45&	291.9468606&	0.125&	58.6  \\ \hline \hline 
\\
   \end{tabular}
    \label{tab:my_label}
\end{table}

\section{tCLEAN robust grid images and image parameters}

\begin{table}[h!]
\begin{center}
\begin{threeparttable}
\centering
\caption{Image properties of the 0.9~mm continuum integrated intensity maps and stacked SO channel maps for a range of Briggs robust parameters}
\label{table:grid}
\begin{tabular}{ccccccc}
0.9~mm Continuum &  & & & & \\
\hline \hline
Robust  & Beam size  & rms (mJy beam$^{-1}$)  & Peak (mJy beam$^{-1}$) & Signal-to-Noise & $\epsilon^{*}$ \\  
\hline
1.0 &   0\farcs27 $\times$ 0\farcs23 (64$^{\circ}$) &  0.019  & 166.1  & 8742 &  0.31 \\  
0.5 &   0\farcs22 $\times$ 0\farcs18 (63$^{\circ}$) &  0.037  & 122.3  & 3305 & 0.56 \\  
0.0 &   0\farcs18 $\times$ 0\farcs14 (66$^{\circ}$)&   0.072 &  82.7  & 1150  & 0.84\\  
-0.5 &   0\farcs15 $\times$ 0\farcs11 (65$^{\circ}$)&  0.117  &  63.0  & 540 &0.92\\  
\hline
& & & & & \\
Stacked SO & & & & & \\
\hline \hline
Robust  & Beam size  & rms$^{+}$ (mJy beam$^{-1}$)  & Peak (mJy beam$^{-1}$) & Signal-to-Noise & $\epsilon^{*}$ \\  
\hline
1.0 &   0\farcs27 $\times$ 0\farcs23 (68$^{\circ}$) &  0.39  & 12.50  & 32.1 &  0.26 \\  
0.5 &   0\farcs23 $\times$ 0\farcs19 (66$^{\circ}$) &  0.69  & 11.07  & 16.0 &0.49 \\  
0.0 &   0\farcs18 $\times$ 0\farcs14 (69$^{\circ}$)&   1.55  &  13.31  & 8.6 &0.75\\  
-0.5 &   0\farcs15 $\times$ 0\farcs11 (65$^{\circ}$)& 2.45  &   13.86  & 5.7 &0.83\\  
\hline
\end{tabular}
\begin{tablenotes}\footnotesize
\item{$^{+}$ With a channel width of 0.12~km~s$^{-1}$ for the line data. $^*$ epsilon from JvM correction (see Section 2 and \citet{2021ApJS..257....2C} for further details on this parameter).}
\end{tablenotes}
\end{threeparttable}
\end{center}
\end{table}

\begin{figure}[h!]
    \centering
    \includegraphics[width=0.88\hsize]{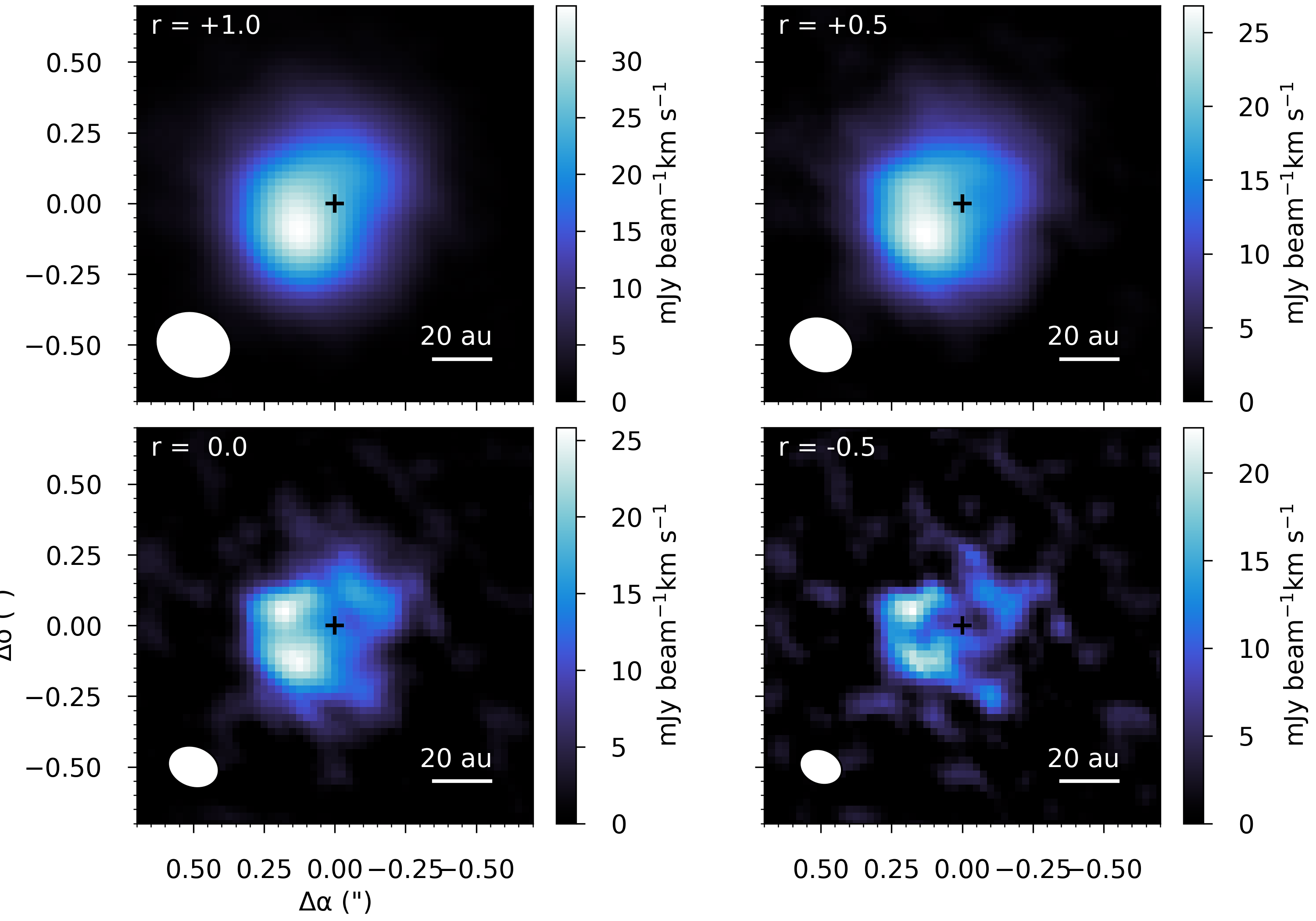}
    \caption{Keplerian masked integrated intensity maps of the stacked SO emission generated over a range of Briggs robust parameters (+1.0, +0.5,0.0, -0.5). The resulting beams are shown in the bottom left hand corner of each panel and the sizes are listed in Table A.1.}
    \label{fig:comparison}
\end{figure}

\newpage

\begin{figure}[h!]
    \centering
    \includegraphics[width=0.88\hsize]{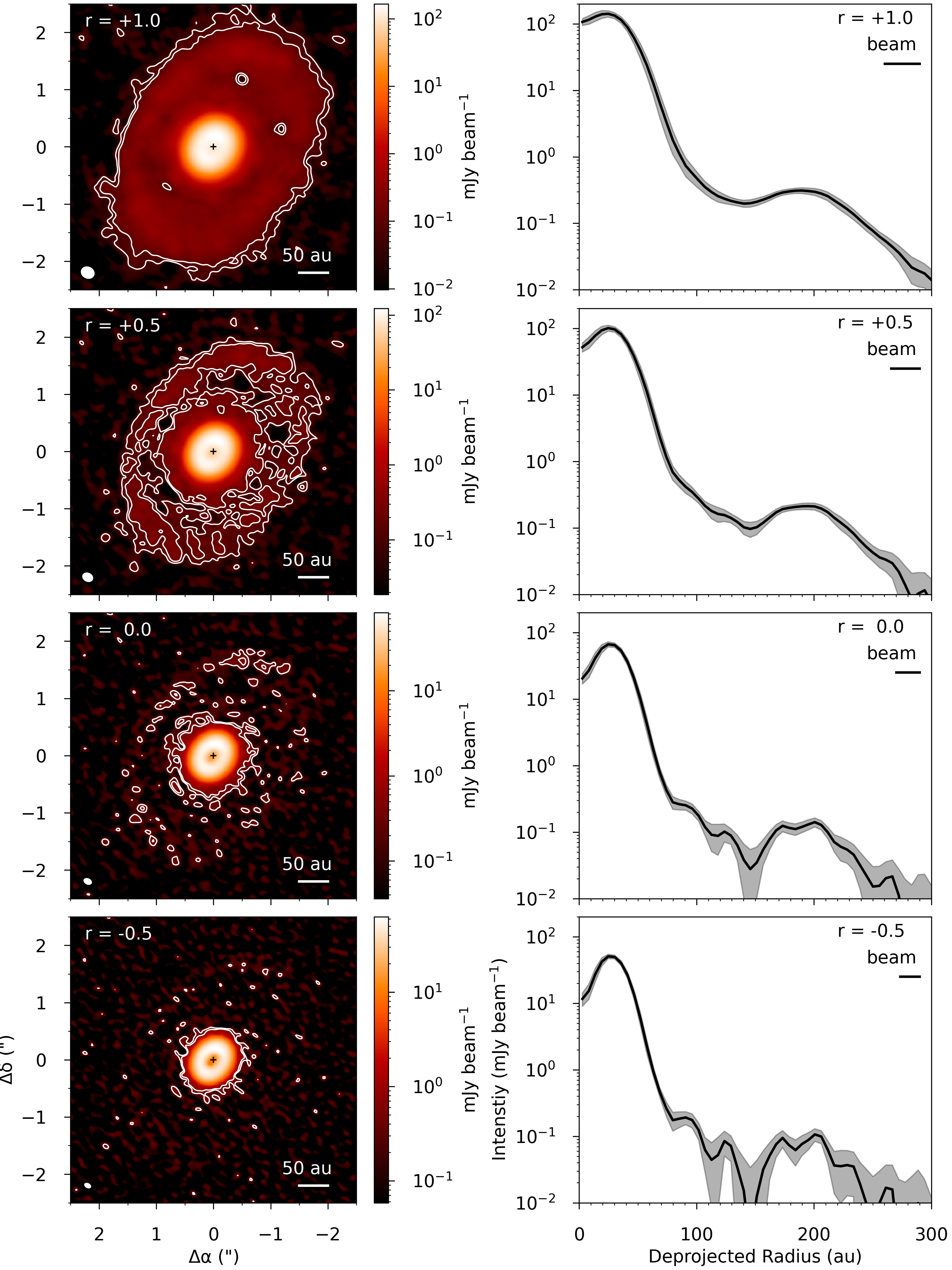}
    \caption{Left: HD~100546 0.9~mm continuum maps with a range of Briggs robust parameters (+1.0, +0.5, 0.0, -0.5). The white contours mark the $\sigma~\times$[3,5] level for each map where $\sigma$ is the rms as listed in Table~B.1. Right: azimuthally averaged and de-projected radial emission profiles for each of the  maps. The error bars are calculated from the standard deviation of the pixel values in each bin divided by the square root of the number of beams per annulus. In the top right corner of each panel the major axis of the beam is represented by a black line.}
    \label{fig:comparison}
\end{figure}

\newpage 

\begin{figure}[h!]
    \centering
    \includegraphics[width=0.95\hsize]{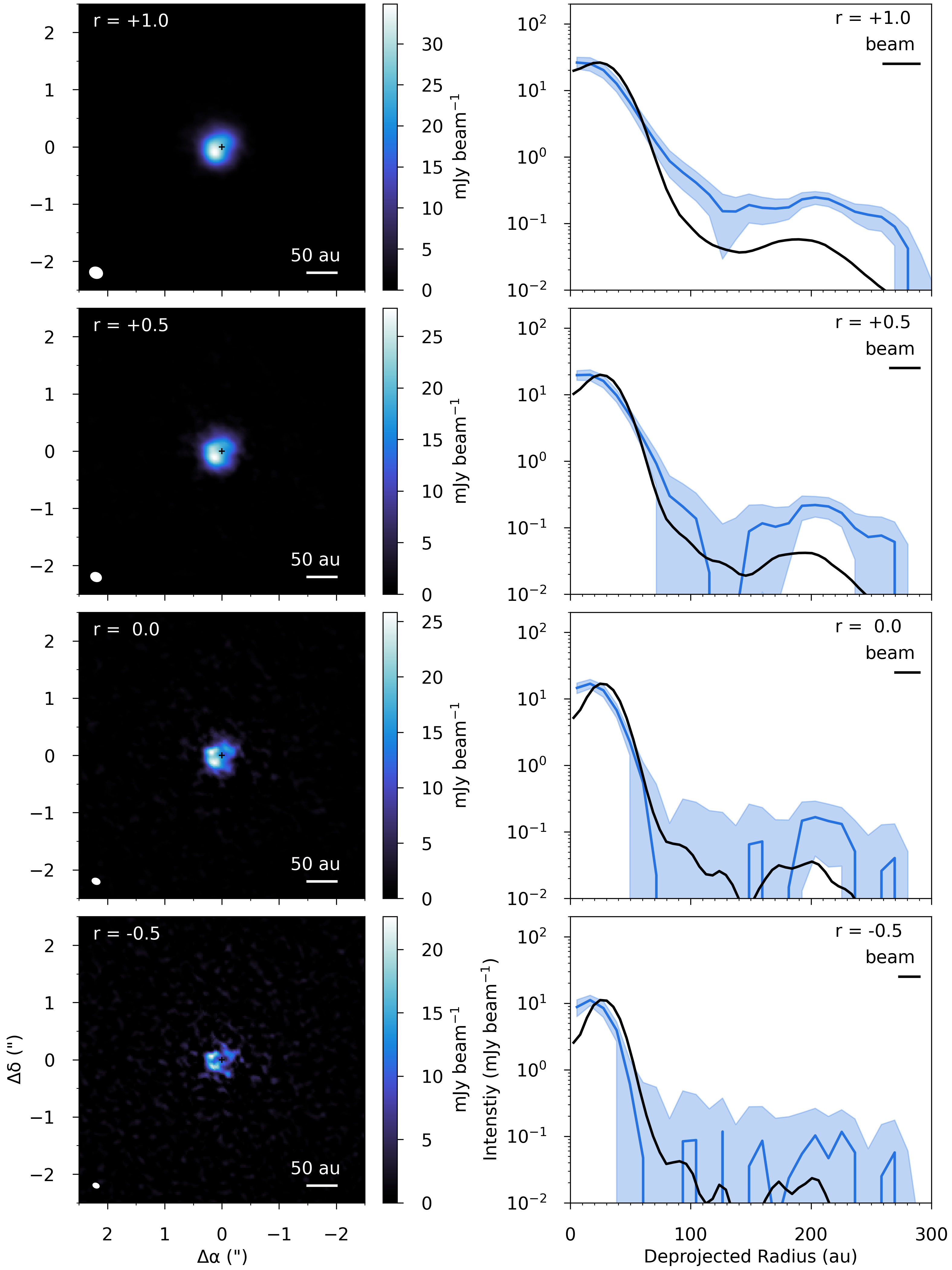}
    \caption{Left: HD~100546 stacked SO Keplerian masked integrated intensity maps with a range of Briggs robust parameters (+1.0, +0.5, 0.0, -0.5). Right: azimuthally averaged and de-projected radial emission profiles for each of the maps with the matching continuum radial profile normalised to the peak in the line radial profile. The error bars are calculated from the standard deviation of the pixel values in each bin divided by the square root of the number of beams per annulus. In the top right corner of each panel the major axis of the beam is represented by a black line.}
    \label{fig:comparison}
\end{figure}

\begin{figure}[h!]
    \centering
    \includegraphics[width=0.45\hsize]{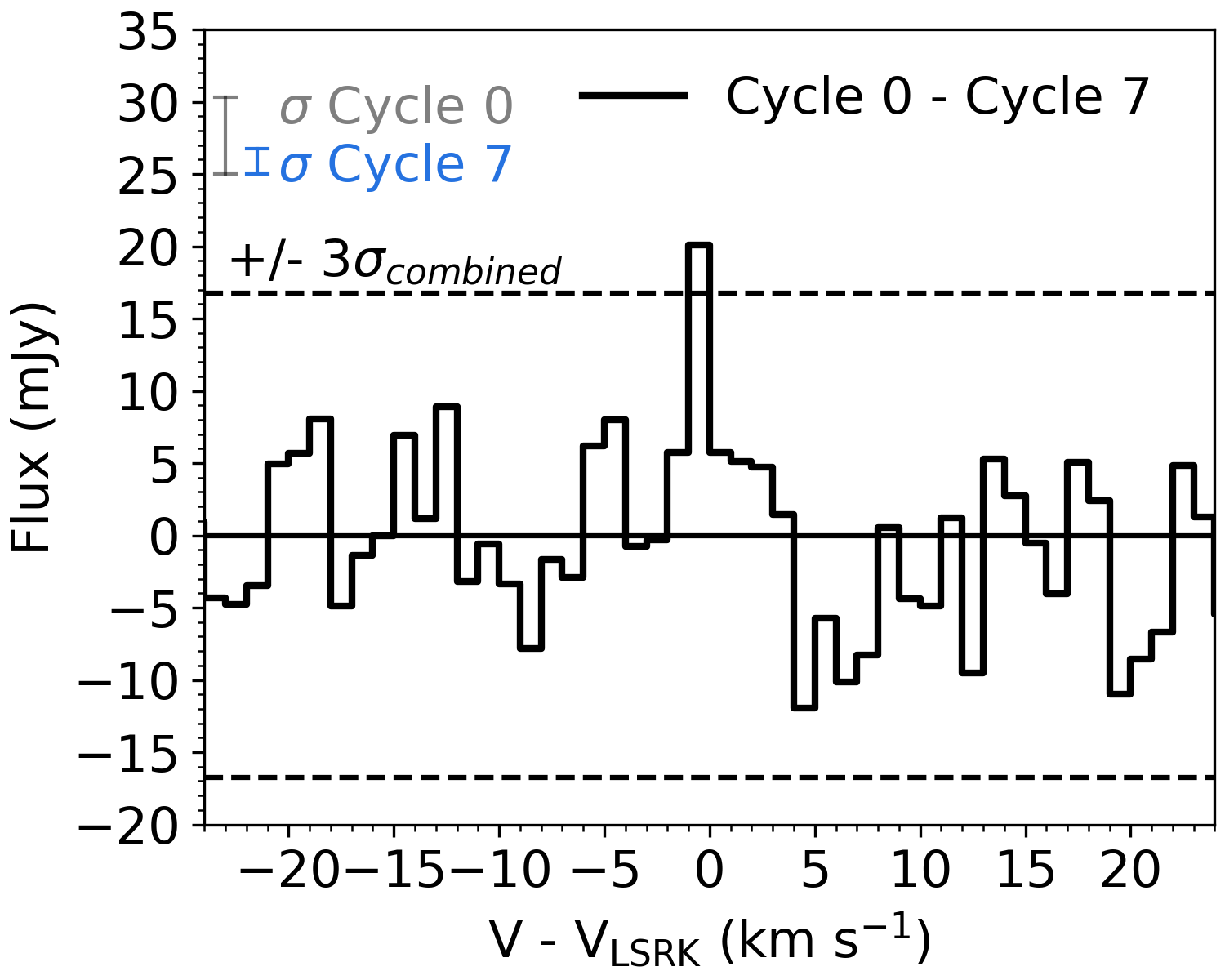}
    \caption{\textcolor{black}{Residual spectra of the Cycle 7 spectra subtracted from the Cycle 0 spectra of the stacked SO emission as shown in Figure 2. 
    The noise calculated from the line free channels in each respective spectrum is shown by a bar in the top left corner of the plot.
    The dashed line shows the $\pm$3~$\sigma$ level where $\sigma_{combined}$ is the rms of each respective spectra added in quadrature.}}
    \label{fig:comparison}
\end{figure}

\newpage

\clearpage

\section{Detection of new sub-millimeter source near HD~100546}

Here we report the detection of a new sub-millimeter source located at a projected separation from HD~100546 of 9\farcs34 with a position angle of 37.7$^{\circ}$ East of North. The peak in continuum emission is at RA:173.360$^{\circ}$ DEC:-70.193$^{\circ}$ and reaches a signal-to-noise of 5 in the continuum image with Briggs robust of +1.0 and has a flux density of 1~mJy. There are a number of previously identified background stars to HD~100546 within our field of view \citep{2001AJ....122.3396G, 2007ApJ...665..512A}. 
The nearest source is field star 9 from Table~3 in \citet{2007ApJ...665..512A} who conclude from the photometry that this is a background star relative to HD~100546 and that due to the reddening, it is within or behind the dark cloud DC296.2-7.9 \citep{1999A&A...345..559V}. Follow up observations of HD~100546 and the surrounding environment with Gemini/NICI confirm their background nature as no significant proper motion of these sources is detected whereas HD~100546 has moved \citep{2013A&A...560A..20B}. 
From GAIA DR3 we investigated the locations of all known sources within our field of view in Figure~B1 and plot those alongside our continuum map and the sources from \citet{2001AJ....122.3396G}.
None of the known sources match with the location of our new sub-mm source. 
The emission we detect may be associated to an envelope/disk around a background star that maybe is young / not optically visible due to both its own envelope/disk and the cloud. As there is no distance measurement to this object any inferred dust mass is highly uncertain. 
\textcolor{black}{
Alternatively, the source we detect could be a background galaxy. We estimate the likelihood of this
using the ALMA 1.1~mm survey from \citet{2016PASJ...68...36H}. 
From this, we would naively expect $\approx$0.1 extragalactic sources of S$>$1~mJy in our primary beam, and so the possibility of this source being a background galaxy may be unlikely, but cannot be discounted.}

\begin{figure}[h!]
    \centering
    \includegraphics[width=0.85\hsize]{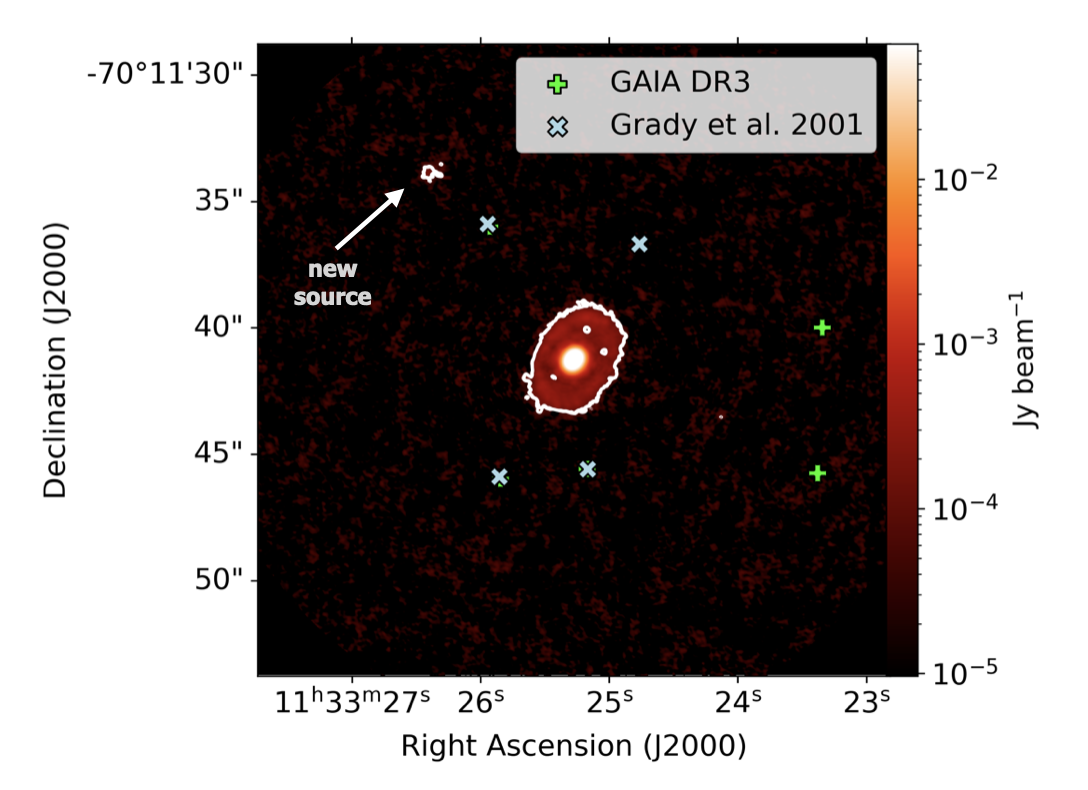}
    \caption{HD~100546 0.9~mm continuum maps with a Briggs robust parameter of +1.0. The white contours marks the $\sigma~\times$[4,5,6] level for each map where $\sigma$ is the rms as listed in Table~A.1. Shown with green crosses are the background stars identified by GAIA DR3 and the blue crosses are from \citet{2001AJ....122.3396G}. }
    \label{fig:comparison}
\end{figure}

\end{appendix}

\end{document}